\documentclass[fleqn,usenatbib]{mnras}
\pdfoutput=1

\usepackage{newtxtext,newtxmath}

\usepackage[T1]{fontenc}
\usepackage{ae,aecompl}

\usepackage{amsmath}
\usepackage{siunitx}
\usepackage{booktabs}
\usepackage[version=4]{mhchem}
\usepackage{hyperref}
\usepackage{pgfplots}
\usepgfplotslibrary{external}
\usetikzlibrary{plotmarks}
\usepgfplotslibrary{colorbrewer}
\pgfplotsset{compat=1.14,
	    cycle list/Dark2-8,
            table/search path={.,ascii},
            spectrum style/.style={
              const plot mark mid,
              no markers,
              extra x ticks       = 0,
              extra x tick labels = ,
              extra y ticks       = 0,
              extra y tick labels = ,
              extra tick style  = { grid = major },
              minor x tick num=4,
              minor y tick num=4,
              xmin=-10, xmax=10,
              xlabel={$v$ [\si{\kilo\m\per\s}]},
              ylabel={$T_\mathrm{mB}$ [\si{\K}]},
            },
            methanol style/.style={
              const plot mark mid,
              no markers,
              extra y ticks       = 0,
              extra y tick labels = ,
              extra tick style  = { grid = major },
              xlabel={$\nu$ [\si{\GHz}]},
              ylabel={$T_\mathrm{mB}$ [\si{\K}]},
              width=\hsize,
              height=9cm,
              xmin=241.69, xmax=241.92
            },
}
\makeatletter
\newlength{\abovecaptionskip}%
\makeatother
\usepackage{threeparttable}

\newcommand{\rh}{r_\mathrm{h}}
\newcommand{\lovejoy}{C/2014 Q2 (Lovejoy)}
\newcommand{\lj}{C/2014 Q2}
\newcommand{\halebopp}{C/1995 O1 (Hale-Bopp)}
\newcommand{\hb}{C/1995 O1}
\newcommand{\rlovejoy}{C/2013 R1 (Lovejoy)}
\newcommand{\rlj}{C/2013 R1}
\newcommand{\lemmon}{C/2012 F6 (Lemmon)}

\newcommand{\herschel}{\textit{Herschel}}
\newcommand{\lime}{\texttt{LIME}}
\newcommand{\cine}{\texttt{CINE}}
\newcommand{\volatiles}{\ce{HCN}, \ce{H2CO}, \ce{CO} and \ce{CH3OH}}
\newcommand{\ulimits}{\ce{CH3CHO} and \ce{NH2CHO}}
\newcommand\rott{\SI{53}{\kelvin}}
\newcommand\rottemp{\SI{53(8)}{\kelvin}}
\newcommand\rtemp{\SI{47(6)}{\kelvin}}
\newcommand\tiram{\SI{67(15)}{\kelvin}}
\newcommand\tkeck{\SI{78(1)}{\kelvin}}
\newcommand\column{\SI[]{1.5(2)e11}{\per\cm\squared}}
\newcommand\columnb{\SI[]{1.5(3)e11}{\per\cm\squared}}
\DeclareSIUnit\mol{mol}
\DeclareSIUnit\mols{\mol\per\s}
\newcommand{\qhcn}{6.1(1)e26}
\newcommand{\qco}{1.2(3)e28}
\newcommand{\qhco}{1.3(2)e27}
\newcommand{\qchoh}{1.3(1)e28}
\newcommand{\qchcho}{<6e26}
\newcommand{\qnhcho}{<1e26}

\newcommand{\hcn}{0.1}
\newcommand{\co}{2.0}
\newcommand{\hco}{0.2}
\newcommand{\choh}{2.2}
\newcommand{\chcho}{<0.1}
\newcommand{\nhcho}{<0.02}

\title[Molecular abundances in comet \lovejoy{}]{Measuring molecular
        abundances in comet \lovejoy{} using the APEX
        telescope\thanks{This publication is based on data acquired with
                the Atacama Pathfinder Experiment (APEX). APEX is a
                collaboration between the Max-Planck-Institut f\"ur
                Radioastronomie, the European Southern Observatory, and
                the Onsala Space Observatory.}
}

\author[de Val-Borro et al.]%
{M.~de~Val-Borro,\textsuperscript{1,2}\thanks{E-mail:
\href{mailto:miguel.devalborro@nasa.gov}{miguel.devalborro@nasa.gov}}
S.~N.~Milam,\textsuperscript{1}
M.~A.~Cordiner,\textsuperscript{1,2}
S.~B.~Charnley,\textsuperscript{1}
I.~M.~Coulson,\textsuperscript{3}
\newauthor
A.~J.~Remijan\textsuperscript{4} and
G.~L.~Villanueva\textsuperscript{1}
\\
\textsuperscript{1}NASA Goddard Space Flight Center, Astrochemistry
Laboratory, 8800 Greenbelt Road, Greenbelt, MD 20771, USA\\
\textsuperscript{2}Department of Physics, Catholic University of
America, Washington, DC 20064, USA\\
\textsuperscript{3}East Asian Observatory, Hilo, HI 96720, USA\\
\textsuperscript{4}National Radio Astronomy Observatory,
Charlottesville, VA 22903, USA
}

\date{Accepted 2017 October 23. Received 2017 October 23; in original form 2017 August 7}

\pubyear{2017}

\begin{document}
\label{firstpage}
\pagerange{\pageref{firstpage}--\pageref{lastpage}}
\maketitle

\begin{abstract}
Comet composition provides critical information on the chemical and
physical processes that took place during the formation of the Solar
system.  We report here on millimetre spectroscopic observations of
the long-period bright comet C/2014 Q2 (Lovejoy) using the Atacama
Pathfinder Experiment (APEX) band 1 receiver between 2015 January UT
16.948 and 18.120, when the comet was at heliocentric distance of \SI{1.30}{au} and
geocentric distance of \SI{0.53}{au}.  Bright comets allow for sensitive
observations of gaseous volatiles that sublimate in their coma.  These
observations allowed us to detect HCN, \ce{CH3OH} (multiple
transitions), \ce{H2CO} and CO, and to measure precise molecular
production rates.  Additionally, sensitive upper limits were derived
on the complex molecules acetaldehyde (\ce{CH3CHO}) and formamide
(\ce{NH2CHO}) based on the average of the strongest lines in the
targeted spectral range to improve the signal-to-noise ratio.
Gas production rates are derived using a
non-LTE molecular excitation calculation involving collisions with
\ce{H2O} and radiative pumping that becomes important in the outer coma
due to solar radiation.  We find a depletion of \ce{CO} in \lovejoy{}
with a production rate relative to water of \SI{\co}{\percent},
and relatively low abundances of $Q(\ce{HCN})/Q(\ce{H2O})$,
\SI{\hcn}{\percent}, and $Q(\ce{H2CO})/Q(\ce{H2O})$,
\SI{\hco}{\percent}.  In contrast the \ce{CH3OH} relative abundance
$Q(\ce{CH3OH})/Q(\ce{H2O})$, \SI{\choh}{\percent}, is close to the mean
value observed in other comets.  The measured production rates are
consistent with values derived for this object from other facilities at
similar wavelengths taking into account the difference in the fields of
view.  Based on the observed mixing ratios of organic molecules in four
bright comets including \lj{}, we find some support for atom addition
reactions on cold dust being the origin of some of the molecules.
\end{abstract}


\section{Introduction}

Comets are small bodies composed of molecular ices and dust particles
that spent most of their lifetime in the outer regions of the Solar
system.  Their nuclei contain pristine material which have not evolved
much since the time of their formation in the early solar nebula.
Therefore, characterizing the chemical composition of the coma can help
to constrain the distribution of molecular material during the epoch of
planet formation.  In addition, studying the role of the volatile ice
composition in the sublimation of material from the surface is important
in the understanding of the processes involved in nucleus activity.
Remote-sensing observations of cometary atmospheres at various
wavelengths are an efficient tool for investigating the physical and
chemical diversity of comets, and substantial efforts have been made in
the last decades to develop a chemical classification of comets
\citep[e.g.][]{1995Icar..118..223A,2004come.book..391B,2011ARA&A..49..471M,2015SSRv..197....9C,2016Icar..278..301D}.
These observations have revealed critical information about the
composition of the primordial material in the solar nebula and the early
formation stages of the Solar system.

Some of the observed molecules appear to be parent species sublimating
isotropically from the nucleus, while others such as HNC and \ce{H2CO} have
been shown to have a distributed source based on spatially resolved
observations using the Atacama Large Millimeter/submillimeter Array
\citep[ALMA;][]{2017ApJ...838..147C}. Possible production mechanisms
for these molecules are gas-phase chemistry in extended coma
or degradation of refractory organic material contained in the nucleus.
To constrain the formation scenarios of daughter molecules using mapping
observations of molecular distributions in cometary comae, a
three-dimensional molecular excitation and radiative transfer model is
required such as the one presented in \citet{2010A&A...523A..25B}.

Comet \lovejoy{} (hereafter referred to as \lj{}) is a long-period comet
with an orbital period \SI{\approx 11000}{yr},
eccentricity of \num{0.998} and inclination \ang{80.3;;}.
The comet was discovered by Terry Lovejoy on 2014 August 17 at a heliocentric distance of
\SI{2.63}{au} \citep{2014CBET.3934....1L}. \lj{} originates from the Oort
cloud and it was one of the most active comets that had a close approach
to the Earth in the last two decades.  The comet was visible to the naked
eye around the time of its perihelion passage on 2015 January UT 30.06
at \SI{1.29}{au} heliocentric distance, reaching apparent magnitude 4.

Thanks to the favourable apparition geometry, 21 different organic
molecules outgassing from the nucleus were detected in \lj{} in the
millimetre domain, including ethyl alcohol, \ce{C2H5OH}, and the
simple sugar glycolaldehyde, \ce{CH2OHCHO} using the Institut de
radioastronomie millim\'etrique (IRAM) \SI{30}{\m} telescope
\citep{2015SciA....115863B}.  The presence of complex organic molecules
suggests that this object originates from the outskirts of the pre-solar nebula.
The detection of HDO emission was accomplished in the millimetre range
of wavelengths using the IRAM \SI{30}{\m} telescope
\citep{2016A&A...589A..78B} and by infrared spectroscopy with the Keck
Observatory \citep{2017ApJ...836L..25P}, increasing the number of known
HDO/\ce{H2O} ratios in comets and confirming their diverse chemical
composition.

Using the Atacama Pathfinder Experiment (APEX) telescope, we have
obtained observations of several molecules to study the volatile
composition in the coma of comet \lj{} near perihelion.  Several
molecules, namely \volatiles{}, were detected and sensitive
upper limits on the abundance of complex molecules \ce{CH3CHO} and
\ce{NH2CHO} were derived.  Using a non-LTE excitation and radiative
transfer code, we computed the production rates for the observed
molecular lines, and compared them with observations from other
facilities and typical ratios measured in comets.

In Section~\ref{sec:observations}, the observations of volatile species
and the reduction method are summarized.  Section~\ref{sec:results}
presents the data analysis of the observations
calculated with a model based on a non-LTE excitation and radiative
transfer code including radiation trapping effects
\citep{2010A&A...523A..25B,2017JOSS.2017..182D} derived from
a previous one-dimensional implementation
\citep{2000A&A...362..697H,2016ascl.soft12009D}.  In
Section~\ref{sec:discussion} we explore different
scenarios for the addition of CO molecules and heavy atoms
on cold interstellar/nebular dust grains to form larger molecules.
Finally, we summarize the results and
discuss the main conclusions in Section~\ref{sec:conclusions}.

\begin{figure*}
  \centering
  \begin{tikzpicture}
    \begin{axis}[spectrum style,name=hcn,
            ymin=-0.5, ymax=2.2,
	    extra x ticks       = ,
            ]
            \addplot[black] table {Q2_HCN_X202_2015-01-16.txt};
       \node at (rel axis cs:0.2,0.9) {\ce{HCN} $J=3$--2};
       \node at (rel axis cs:0.78,0.9) {UT 16.948 Jan 2015};
       \draw[green,thin] (-2.3545,-0.2) -- (-2.3545,0.08680031-0.2);
       \draw[green,thin] (-0.6149,-0.2) -- (-0.6149,0.00247983-0.2);
       \draw[green,thin] (-0.0744,-0.2) -- (-0.0744, 1.0042024-0.2);
       \draw[green,thin] (0,-0.2) -- (0,0.69425858-0.2);
       \draw[green,thin] (0.2766,-0.2) -- (0.2766,0.46862047-0.2);
       \draw[green,thin] (1.7394,-0.2) -- (1.7394,0.08680031-0.2);
       \draw[green,thin] (1.7394,-0.2) -- (-2.3545,-0.2);
    \end{axis}
    \begin{axis}[spectrum style,
      at={(hcn.east)},anchor=west,
      yticklabel=\empty, ylabel=\empty,
      xtick={-5,0,5,10},
      minor xtick={-9,-8,...,9},
      ymin=-0.5, ymax=2.2,
      extra x ticks       = ,
    ]
      \addplot[black] table {Q2_HCN_X202_2015-01-18.txt};
      \node at (rel axis cs:0.2,0.9) {\ce{HCN} $J=3$--2};
      \node at (rel axis cs:0.78,0.9) {UT 17.989 Jan 2015};
      \draw[green,thin] (-2.3545,-0.2) -- (-2.3545,0.08680031-0.2);
      \draw[green,thin] (-0.6149,-0.2) -- (-0.6149,0.00247983-0.2);
      \draw[green,thin] (-0.0744,-0.2) -- (-0.0744, 1.0042024-0.2);
      \draw[green,thin] (0,-0.2) -- (0,0.69425858-0.2);
      \draw[green,thin] (0.2766,-0.2) -- (0.2766,0.46862047-0.2);
      \draw[green,thin] (1.7394,-0.2) -- (1.7394,0.08680031-0.2);
      \draw[green,thin] (1.7394,-0.2) -- (-2.3545,-0.2);
    \end{axis}
  \end{tikzpicture}
  \caption{\ce{HCN} J=3--2 line observed in the spectra of
	  comet \lovejoy{} with the XFFTS2 backend plotted in the
	  cometocentric rest frame. The unresolved HCN hyperfine
	  transitions are shown by the vertical lines.
  }
  \label{fig:hcn}
\end{figure*}
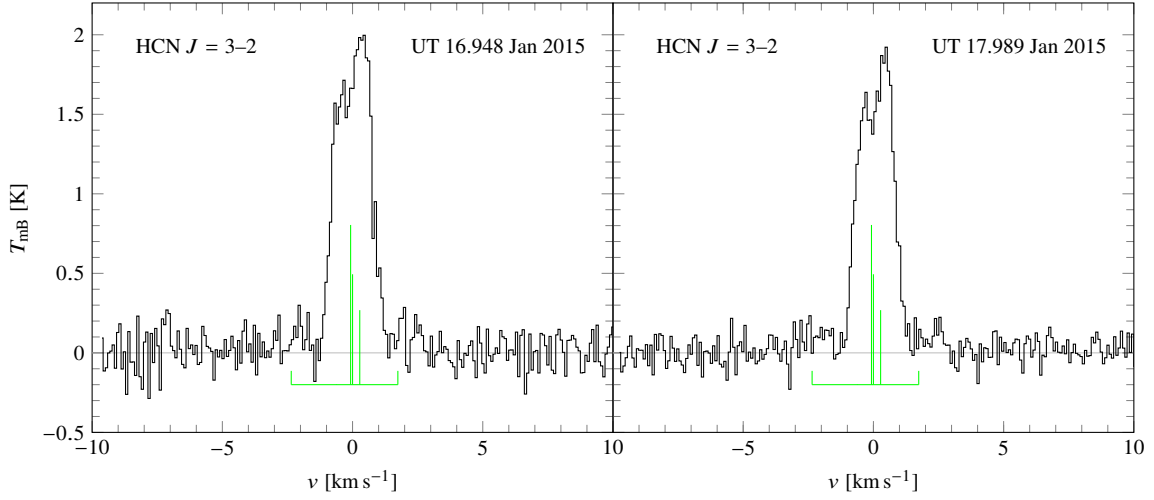

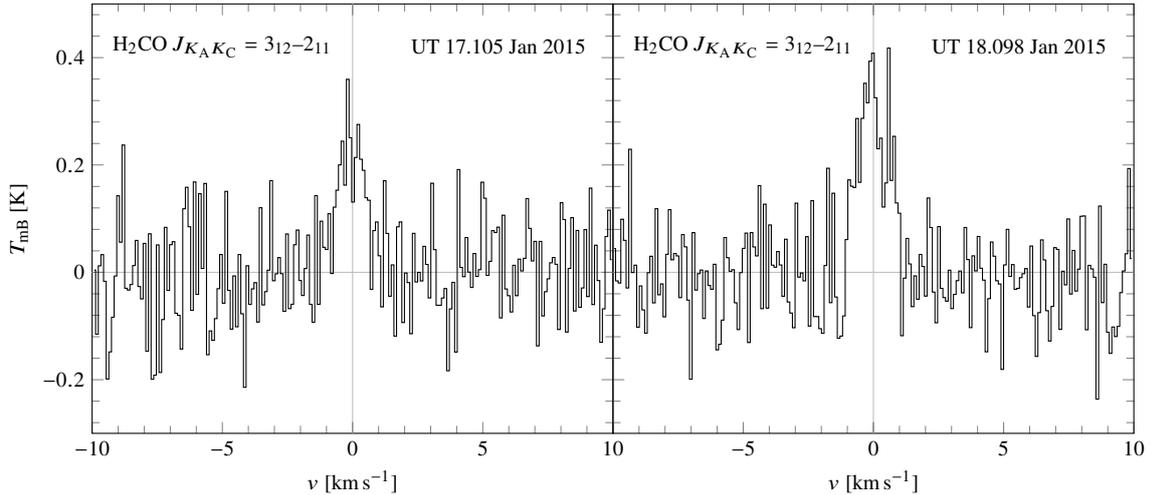
\begin{figure*}
  \centering
  \begin{tikzpicture}
    \begin{axis}[spectrum style,name=h2co,
            ymin=-0.3, ymax=0.5]
      \addplot[black] table {Q2_H2CO_X202_2015-01-17.txt};
      \node at (rel axis cs:0.25,0.9) {\ce{H2CO} $J_{K_\mathrm{A}K_\mathrm{C}} = 3_{12}$--$2_{11}$};
      \node at (rel axis cs:0.78,0.9) {UT 17.105 Jan 2015};
    \end{axis}
    \begin{axis}[spectrum style,
      at={(h2co.east)},anchor=west,
      yticklabel=\empty, ylabel=\empty,
      xtick={-5,0,5,10},
      minor xtick={-9,-8,...,9},
      ymin=-0.3, ymax=0.5]
      \addplot[black] table {Q2_H2CO_X202_2015-01-18.txt};
      \node at (rel axis cs:0.25,0.9) {\ce{H2CO} $J_{K_\mathrm{A}K_\mathrm{C}} = 3_{12}$--$2_{11}$};
      \node at (rel axis cs:0.78,0.9) {UT 18.098 Jan 2015};
    \end{axis}
  \end{tikzpicture}
  \caption{\ce{H2CO} $\mathrm{J}_{K_\mathrm{A}K_\mathrm{C}} = 3_{12}$--$2_{11}$
    line observed in the spectra of comet \lovejoy{} with the XFFTS2
    backend plotted in the cometocentric rest frame.
  }
  \label{fig:h2co}
\end{figure*}

\section{Observations}
\label{sec:observations}

We carried out high resolution observations of comet \lj{} using the
\SI{12}{\m} single-dish APEX telescope located at
\SI{5100}{\m} above sea level in the Atacama desert in northern Chile
\citep{2006A&A...454L..13G}. Our observing program was executed
on two nights from UT 16.948 to 18.120 January 2015.
The Swedish Heterodyne
Facility Instrument \citep[SHeFI;][]{2008A&A...490.1157V} consists of
four single-band superconductor--insulator--superconductor (SIS)
heterodyne receivers equipped with several backends. We have used the APEX-1 receiver that operates in
the \SIrange{213}{275}{\GHz} band for our observations in combination
with the eXtended bandwidth Fast Fourier Transform Spectrometer (XFFTS)
backend. XFFTS offers a high instantaneous bandwidth of \SI{2.5}{\GHz}
and \num{32768} channels with a \SI{76.29}{\kHz} spectral resolution.
The full width at
half-maximum (FWHM) of the APEX beam ranges from
\SIrange{23}{27}{\arcsecond}
at the considered frequencies, which correspond to distances of
\SIrange{10000}{12000}{\kilo\m} at the distance of the comet.
The line intensities for a given frequency and each of the backend
groups in the XFFTS are calibrated with the recommended factors.

Observing conditions were favourable throughout the observing period.
The precipitable water vapour over the interval of the observations
remained between \SIrange{3.0}{3.7}{\milli\m} on the night of 2015
January 16 and between \SIrange{3.2}{4.0}{\milli\m} on 2015 January 17.
Pointing and focus calibration observations were carried out on Mars and
Jupiter because they were close to the comet during the course of our
observations.  We obtained flux density reference observations of bright
standard sources for calibration purposes alternated with the comet
observations on a regular basis.  The observed sources were the Mira
variables o Ceti, IK Tauri and R Doradus, the evolved carbon stars IRC
+10216 and R Leporis, and the Orion Molecular Cloud Complex.  We used
the latest ephemeris provided by JPL's HORIZONS Solar system
service\footnote{\url{http://ssd.jpl.nasa.gov/?horizons}} during the
observations to track the position and relative motion of the comet with
respect to the observer \citep{1996DPS....28.2504G}.

Due to the stability of the instrument and the good observing
conditions at the APEX site, the calibration of the spectra
did not pose special problems.  We used the open-source
\texttt{pyspeckit} spectroscopic package to read the data files
downloaded from the APEX archive \citep{2011ascl.soft09001G}.
\texttt{pyspeckit} supports several file formats including recent
versions of the CLASS file format.

Since APEX observations are provided in the Local Standard of Rest
Kinematic (LSRK), frequency calibration is required to convert to the
cometocentric frame in order to find out the Doppler shift of the observed
lines with respect to the rest frequency.
Converting velocities from the LSRK frame to the geocentric frame
is an operation that depends on the position of the comet and
the time of the observation.
We adopted the value for the Standard Solar Motion as that used by
the APEX staff: \SI{20}{\km\per\s} towards $\alpha_{1900} =
\ang{18;3;50.24}$, $\delta_{1900} = \ang{30;0;16.8}$ (D. Muders, private
communication).
Finally the position and velocity of the comet were calculated for each
scan with \SI{\approx 30}{\s} integration using the
\texttt{callhorizons} code to access the JPL HORIZONS ephemerides of
comet \lj{}\footnote{The source code is available at
\url{https://github.com/mommermi/callhorizons}}.

\begin{table*}
\begin{threeparttable}
  \centering
  \caption{Observing log of comet \lovejoy{} obtained with APEX.
  }
  \begin{tabular}{ccc
		  S[table-format = 2.1,
		  output-decimal-marker = {\fdg}]
		  c
		  S[table-format = 2.1,
		    output-decimal-marker = {\fm}]
		  S[table-format = 2.1,
		    output-decimal-marker = {\farcs}]
		  }
    \toprule
    Date (UT)                & $\rh$\tnote{\emph{a}}  & $\Delta$\tnote{\emph{b}}  &
    {$\phi$\tnote{\emph{c}}} & Species                &
    {Int.\tnote{\emph{d}}} & {Beam FWHM}              \\
    mm/dd.ddd                & (au)                   & (au)                      &
    {(\si{\degree})}         &                        &
    {(\si{\minute})}       & {(\si{\arcsecond})}      \\
    \midrule
    1/16.948&1.3055&0.5281&42.56& \ce{HCN} &4.5&23.5\\
1/17.002&1.3054&0.5287&42.61& \ce{CH3OH} &22.5&25.8\\
1/17.078&1.3052&0.5296&42.68& \ce{CO} &12.4&27.1\\
1/17.105&1.3052&0.5299&42.71& \ce{H2CO} &3.5&27.6\\
1/17.115&1.3051&0.5300&42.72& \ce{CH3OH} &2.2&25.8\\
1/17.989&1.3032&0.5406&43.46& \ce{HCN} &7.1&23.5\\
1/18.030&1.3031&0.5411&43.50& \ce{CH3OH} &15.0&25.8\\
1/18.075&1.3030&0.5417&43.54& \ce{CO} &10.6&27.1\\
1/18.098&1.3030&0.5420&43.55& \ce{H2CO} &4.4&27.6\\
1/18.110&1.3030&0.5421&43.56& \ce{CH3OH} &2.6&25.8\\
1/18.120&1.3029&0.5423&43.57& \ce{HCN} &3.1&23.5\\

    \bottomrule
  \end{tabular}
  \begin{tablenotes}
    \item [\emph{a}] Heliocentric distance.
    \item [\emph{b}] Geocentric distance.
    \item [\emph{c}] Solar phase angle.
    \item [\emph{d}] Total integration time.
  \end{tablenotes}
  \label{tab:log}
\end{threeparttable}
\end{table*}

We present the observing circumstances and total integration times for
each molecule in Table~\ref{tab:log}.
We detected four molecules (\volatiles{}) and obtained significant upper
limits on the emission of the complex molecules acetaldehyde
(\ce{CH3CHO}) and formamide (\ce{NH2CHO}) by averaging several lines
shown in Table~\ref{tab:nh2cho}.

\begin{table}
  \centering
  \caption{Targeted \ce{NH2CHO} and \ce{CH3CHO} lines in comet
	  \lovejoy{}. The rest-frame line frequencies are obtained from
	  the JPL and CDMS catalogues.
  }
  \label{tab:nh2cho}
  \begin{tabular}{c
		  S[table-format = 3.7]
		  cc
           }
    \toprule
    Molecule & {Frequency} & Upper & Lower \\
       & {(\si{\GHz})} & & \\
    \midrule
\ce{NH2CHO} & 227.6061762  & $11_{0, 11}$ & $10_{0, 10}$ \\
\ce{NH2CHO} & 232.2742626  & $11_{2, 10}$ & $10_{2,  9}$ \\
\ce{NH2CHO} & 239.9523537  & $11_{1, 10}$ & $10_{1,  9}$ \\
\ce{NH2CHO} & 243.5216888  & $12_{1, 12}$ & $11_{1, 11}$ \\
\ce{NH2CHO} & 267.0633926  & $13_{0, 13}$ & $12_{0, 12}$ \\
%
\ce{CH3CHO} & 226.5516216  & $12_{0, 12}$ & $11_{0, 11}$ \\
\ce{CH3CHO} & 226.5927251  & $12_{0, 12}$ & $11_{0, 11}$ \\
\ce{CH3CHO} & 230.3018800  & $12_{2, 11}$ & $11_{2, 10}$ \\
\ce{CH3CHO} & 230.3157400  & $12_{2, 11}$ & $11_{2, 10}$ \\
\ce{CH3CHO} & 231.5952727  & $12_{3,10}$ & $11_{3,9}$ \\
\ce{CH3CHO} & 231.7487186  & $12_{3,10}$ & $11_{3,9}$ \\
\ce{CH3CHO} & 231.8475795  & $12_{3,9 }$ & $11_{3,8}$ \\
\ce{CH3CHO} & 231.9683853  & $12_{3,9 }$ & $11_{3,8}$ \\
\ce{CH3CHO} & 242.1060180  & $13_{1, 13}$ & $12_{1, 12}$ \\
\ce{CH3CHO} & 242.1181400  & $13_{1, 13}$ & $12_{1, 12}$ \\

    \bottomrule
  \end{tabular}
\end{table}

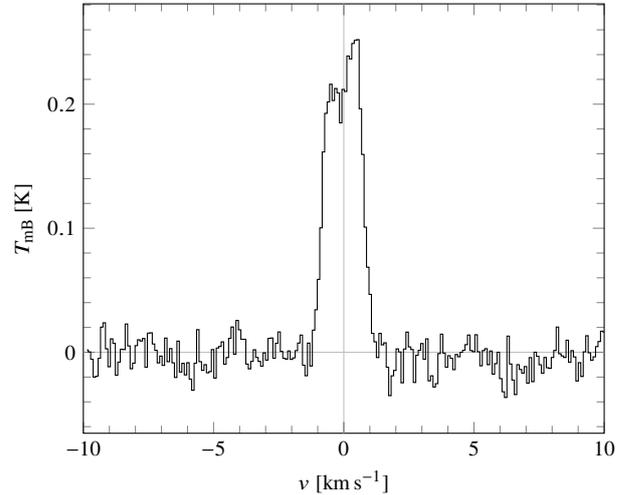
\begin{figure}
  \centering
  \begin{tikzpicture}
    \begin{axis}[spectrum style]
      \addplot[black] table {ch3oh_av.txt};
    \end{axis}
  \end{tikzpicture}
  \caption{Averaged spectrum of \ce{CH3OH} emission lines
	  with frequencies
	  \SIlist{241.700;241.767;241.791;241.879;241.888}{\GHz}
	  in comet \lovejoy{} obtained
	  with the XFFTS2 backend.
  }
  \label{fig:ch3oh_av}
\end{figure}

\begin{figure*}
  \begin{tikzpicture}
    \begin{axis}[spectrum style,name=h2co,
            ymin=-0.1, ymax=0.12]
      \addplot[black] table {Q2_CO_X202_2015-01-17.txt};
      \node at (rel axis cs:0.25,0.9) {\ce{CO} $J = 2$--1};
      \node at (rel axis cs:0.78,0.9) {UT 17.078 Jan 2015};
    \end{axis}
    \begin{axis}[spectrum style,
      at={(h2co.east)},anchor=west,
      yticklabel=\empty, ylabel=\empty,
      xtick={-5,0,5,10},
      minor xtick={-9,-8,...,9},
      ymin=-0.1, ymax=0.12]
      \addplot[black] table {Q2_CO_X202_2015-01-18.txt};
      \node at (rel axis cs:0.25,0.9) {\ce{CO} $J = 2$--1};
      \node at (rel axis cs:0.78,0.9) {UT 18.075 Jan 2015};
    \end{axis}
  \end{tikzpicture}
  \caption{\ce{CO}-averaged spectrum in comet \lovejoy{} obtained
	  with a \SI{12.3}{\minute} (left-hand pannel) and \SI{12.6}{\minute}
	  (right-hand pannel) integration time using the XFFTS2 receiver.
  }
  \label{fig:co}
\end{figure*}
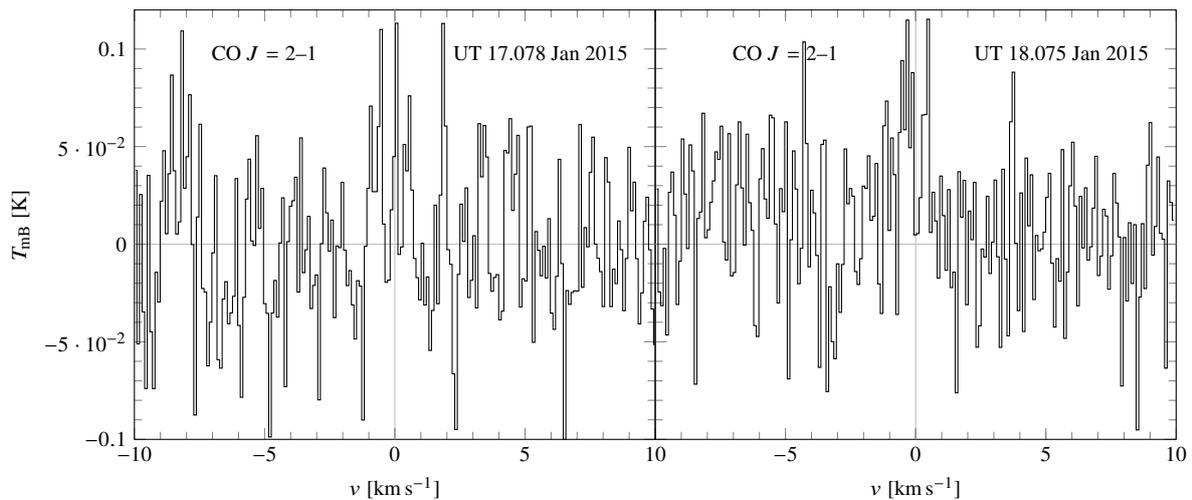

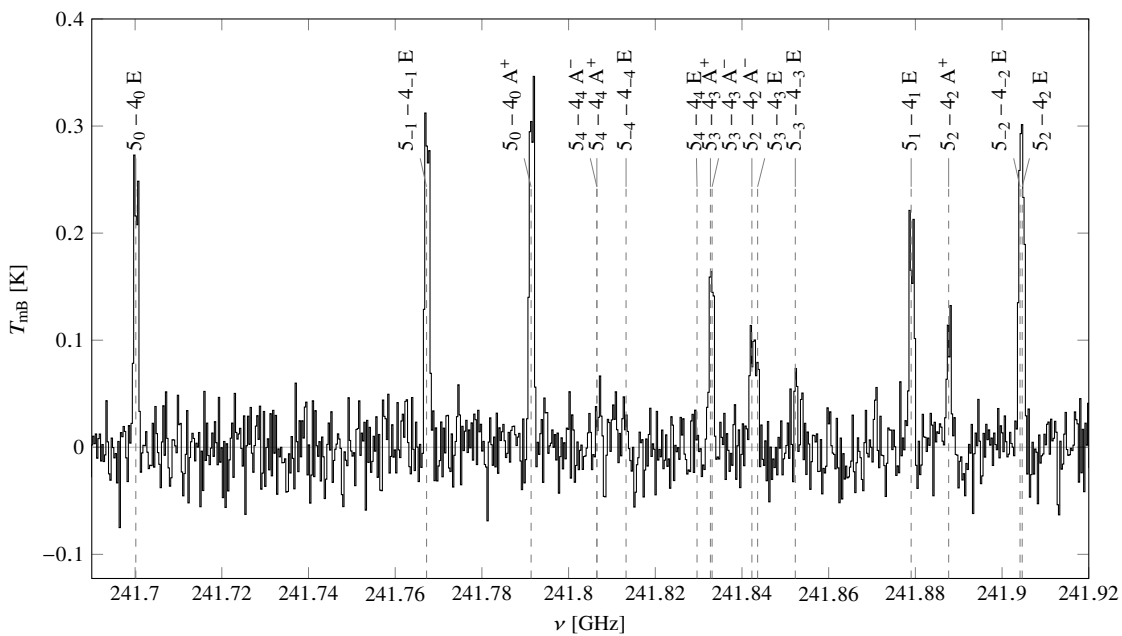
\begin{figure*}
  \centering
  \begin{tikzpicture}
    \begin{axis}[methanol style, ymax=0.4]
      \addplot[black] table {Q2_CH3OH_X202_2015-01-18_f.txt};
      \foreach \freq/\angle/\trans in {
      241.7001590/right/$5_0$ -- $4_0$ E,
      241.7672340/5/$5_{-1}$ -- $4_{-1}$ E,
      241.7913520/5/$5_0$ -- $4_0$ A$^+$,
      241.8065240/5/$5_4$ -- $4_4$ A$^-$,
      241.8065250/right/$5_4$ -- $4_4$ A$^+$,
      241.8132550/right/$5_{-4}$ -- $4_{-4}$ E,
      241.8296290/3/$5_4$ -- $4_4$ E,
      241.8327180/right/$5_3$ -- $4_3$ A$^+$,
      241.8331060/-4/$5_3$ -- $4_3$ A$^-$,
      241.8422840/right/$5_2$ -- $4_2$ A$^-$,
      241.8436040/-4/$5_3$ -- $4_3$ E,
      241.8522990/right/$5_{-3}$ -- $4_{-3}$ E,
      241.8790250/right/$5_1$ -- $4_1$ E,
      241.8876740/right/$5_2$ -- $4_2$ A$^+$,
      241.9041470/5/$5_{-2}$ -- $4_{-2}$ E,
      241.9046430/-5/$5_2$ -- $4_2$ E
      }
      { \edef\temp{\noexpand
        \draw[gray,dashed]
          ({axis cs:\freq,0}|-{rel axis cs:0,0}) --
          ({axis cs:\freq,0}|-{rel axis cs:0,0.7});
        }
        \temp
        \edef\temp{\noexpand
        \node[
          coordinate,
          pin = {[rotate=90]\angle:{\trans}}
        ] at ({axis cs:\freq,0}|-{rel axis cs:0,0.7}) {};
        }
        \temp
      }
    \end{axis}
  \end{tikzpicture}
  \caption{Spectrum of \ce{CH3OH} emission in comet \lovejoy{} obtained on
    2015 January UT 18.042 with a \SI{17.6}{\minute} total integration
    time using the XFFTS2 backend.  The spectrum was resampled to a
    \SI{300}{\kHz} resolution per
    channel using a rectangular window function to increase the S/N
    ratio.  \ce{CH3OH} transitions are indicated by the dashed lines.
  }
  \label{fig:ch3oh}
\end{figure*}

Fig.~\ref{fig:hcn} shows the \ce{HCN} J = 3--2 transition
observed in two different epochs in comet \lj{} after baseline removal.
For this species, we used the line frequencies for the hyperfine
structure components from the latest version of the Cologne Database for Molecular
Spectroscopy \citep[CDMS;][]{2016JMoSp.327...95E}.
We show the \ce{H2CO} $\mathrm{J}_{K_\mathrm{A}K_\mathrm{C}} = 3_{12}$--$2_{11}$
transition in Fig.~\ref{fig:h2co}.  A marginal detection is obtained
for the CO (J = 2--1) line in both the backend group 1 (4.6$\sigma$)
and the backend group 2 (5.2$\sigma$) only on the second night of
observations (January UT 18.075).
In Fig.~\ref{fig:ch3oh_av} we show an
average of several of the individually resolved methanol lines that are
observed simultaneously in one of the receiver settings.  
We show these CO spectra in
Fig.~\ref{fig:co}.  Several methanol transitions were observed
simultaneously in the E-\ce{CH3OH} and A-\ce{CH3OH} ladders of levels
with J = 5, including some that are blended with other transitions.
Fig.~\ref{fig:ch3oh} shows the \ce{CH3OH} spectrum from
January UT  18.042.

The HCN line is detected with high-signal-to-noise ratio (S/N) and is
the only detected transition that is expected to be optically thick.
However, the observed line asymmetry could be caused by either
line-opacity effects in the foreground of the coma or anisotropic
outgassing.
Since emission lines from other molecules do not have enough
S/N to resolve the line profile, we have averaged several of the
methanol lines.
The averaged
line shows a significant redshift and asymmetry
(Fig.~\ref{fig:ch3oh_av}) that seems similarly
constructed as in the HCN J = 3--2 transition. The \ce{CH3OH} emission
is predicted to be optically thin.  Thus the similarity in the \ce{HCN}
and \ce{CH3OH} line shapes could be more likely explained by an asymmetric
outgassing with preferential emanation from active regions in the
nucleus.

Table~\ref{tab:apex} shows the line intensity for the transitions
detected with 1$\sigma$ uncertainties as the average value of the
intensities measured with the XFFTS1 and XFFTS2 backend.
Since we did not measure a substantial variation in the production rates
of the observed molecules, we have
co-added observations of HCN and \ce{CH3OH} obtained on the same night
to increase the S/N of the line emission.  Line areas were obtained by
numerically integrating the signal over the fitted baseline and the
statistical uncertainty in the integrated line intensity was calculated
using the root mean square (rms) noise in the background after
subtracting the fitted baseline.
There is a good agreement between the intensities measured by the XFFTS1
and XFFTS2 backends with a variation \SI{< 10}{\percent} for detections
with S/N \num{>5}.
Line frequencies listed in
Table~\ref{tab:apex} were obtained from the latest online edition of the
JPL molecular spectroscopy catalogue \citep{1998JQSRT..60..883P}.

\begin{table*}
\begin{threeparttable}
  \centering
  \caption{Observed line intensities for \volatiles{} integrated over velocity
    and rms error in comet \lovejoy{} measured as the
    average value of the intensities from the XFFTS1 and XFFTS2
    backends between 2015 January UT 16.948--18.120.  Statistical
    uncertainties are shown for the integrated line intensity.
  }
  \label{tab:apex}
  \begin{tabular}{c
		  c
		  S[table-format = 3.5]
		  ccc
		  S[table-format = 2.3,
		    output-decimal-marker = {\fm}]
		  S[table-format = 1.3]
		  S[table-format = 1.3(2)]
		  S[table-format = 2.1]
           }
    \toprule
    Molecule & Form & {Frequency} & Upper & Lower & Date (UT) & {Exposure} &
       $\sigma_{T_\mathrm{mB}}$ & ${\int T_\mathrm{mB}\, dv}$ & {Variation\tnote{\emph{a}}}
       \\
       & & {(\si{\GHz})} & & & mm/dd.ddd & {(\si{\minute})}
       & {(\si{\kelvin})} & {(\si{\kelvin\km\per\s})} & {(\si{\percent})}
       \\
    \midrule
      \ce{HCN} &&265.88643&$3{}$&$2{}$&1/16.948&4.461&0.096& 3.157\pm0.056 &0.4\\
 \ce{CH3OH} &E&241.70016&$5_{0}$&$4_{0}$&1/17.012&24.698&0.045& 0.414\pm0.027 &5.3\\
 \ce{CH3OH} &E&241.76723&$5_{-1}$&$4_{-1}$&1/17.012&24.698&0.045& 0.497\pm0.028 &6.6\\
 \ce{CH3OH} &A&241.79135&$5_{0}$&$4_{0}$&1/17.012&24.698&0.045& 0.508\pm0.027 &2.0\\
 \ce{CH3OH} &E&241.85230&$5_{-3}$&$4_{-3}$&1/17.012&24.698&0.045& 0.068\pm0.024 &8.3\\
 \ce{CH3OH} &E&241.87903&$5_{1}$&$4_{1}$&1/17.012&24.698&0.045& 0.312\pm0.025 &5.0\\
 \ce{CH3OH} &A&241.88767&$5_{2}$&$4_{2}$&1/17.012&24.698&0.045& 0.216\pm0.026 &1.7\\
 \ce{CO} &&230.53800&$2{}$&$1{}$&1/17.078&12.371&0.039& 0.014\pm0.021 &48.9\\
 \ce{H2CO} &&225.69778&$3_{12}$&$2_{11}$&1/17.105&3.522&0.085& 0.311\pm0.046 &12.4\\
 \ce{HCN} &&265.88643&$3{}$&$2{}$&1/18.029&10.155&0.068& 2.839\pm0.039 &1.3\\
 \ce{CH3OH} &E&241.70016&$5_{0}$&$4_{0}$&1/18.042&17.625&0.050& 0.409\pm0.030 &0.7\\
 \ce{CH3OH} &E&241.76723&$5_{-1}$&$4_{-1}$&1/18.042&17.625&0.050& 0.510\pm0.030 &2.7\\
 \ce{CH3OH} &A&241.79135&$5_{0}$&$4_{0}$&1/18.042&17.625&0.050& 0.531\pm0.030 &0.1\\
 \ce{CH3OH} &E&241.85230&$5_{-3}$&$4_{-3}$&1/18.042&17.625&0.050& 0.117\pm0.037 &3.3\\
 \ce{CH3OH} &E&241.87903&$5_{1}$&$4_{1}$&1/18.042&17.625&0.050& 0.346\pm0.031 &4.9\\
 \ce{CH3OH} &A&241.88767&$5_{2}$&$4_{2}$&1/18.042&17.625&0.050& 0.158\pm0.029 &16.6\\
 \ce{CO} &&230.53800&$2{}$&$1{}$&1/18.075&10.566&0.040& 0.123\pm0.029 &19.7\\
 \ce{H2CO} &&225.69778&$3_{12}$&$2_{11}$&1/18.098&4.403&0.073& 0.467\pm0.043 &3.7\\

    \bottomrule
  \end{tabular}
  \begin{tablenotes}
    \item [\emph{a}] Variation in line intensity between the two XFFTS backends.
  \end{tablenotes}
\end{threeparttable}
\end{table*}

\section{Results}
\label{sec:results}

\subsection{Rotational temperatures}
\label{sec:rottemp}

To derive an estimation of the coma kinetic temperature in our radiative
transfer model, we calculate the \ce{CH3OH} rotational temperature
obtained in two different dates.  Methanol rotational transitions appear
in multiple lines at millimetre wavelengths that are well suited to estimate
the rotational temperature and excitation conditions in the coma in the
optically thin limit.  The rotational levels of \ce{CH3OH} listed in
Table~\ref{tab:apex} are described with three quantum numbers ($J_K\
T_\mathrm{s}$) following the notation by \citet{1999JMoSp.194..171M},
where J is the total angular momentum, $K$ its projection along the
symmetry axis, and $T_\mathrm{s}$ is the torsional symmetry state
(A\textsuperscript{+}, A\textsuperscript{-}, E$_1$ or E$_2$),
corresponding to different configurations of the nuclear spin states of
the \ce{CH3} group.  The difference between the E$_1$ and E$_2$ states is
indicated by the sign of the quantum number $K$ with a positive sign
corresponding to E$_1$ levels and a negative sign to E$_2$ levels.
We have observed several \ce{CH3OH} transitions around \SI{242}{\GHz}
that sample rotational levels with quantum number $J = 5$.

Assuming that the population distribution
of the levels sampled by the emission lines is in local thermodynamical
equilibrium (LTE), or described by a Boltzmann distribution
characterized by a single temperature,  the column density of the upper
transition level within the beam, $\langle N_\mathrm{u} \rangle$, can be
expressed as
\begin{equation}
  \langle N_\mathrm{u} \rangle =  \langle N \rangle
  \frac{g_\mathrm{u}}{Z(T_\mathrm{rot})}
  \exp\left(-\frac{E_\mathrm{u}}{k_\mathrm{B}T_\mathrm{rot}}\right),
\end{equation}
where $g_\mathrm{u}$ is the degeneracy of the upper level, $Z$ denotes
the rotational partition function, which is a function of temperature,
$T_\mathrm{rot}$ is the rotational temperature, $E_\mathrm{u}$ is the
energy of the upper state, $k_\mathrm{B}$ represents the Boltzmann
constant and $\langle N \rangle$ is the total column density averaged
over the beam.  
We use the rotational diagram technique described in detail in
\citet{1994A&A...287..647B}.  
This methods allows us to determine the rotational
temperature, $T_\mathrm{rot}$, that describes the relative population of
the \ce{CH3OH} upper states of the observed transitions, by fitting all
the transitions detected above the 3$\sigma$ detection limit with both
backends.

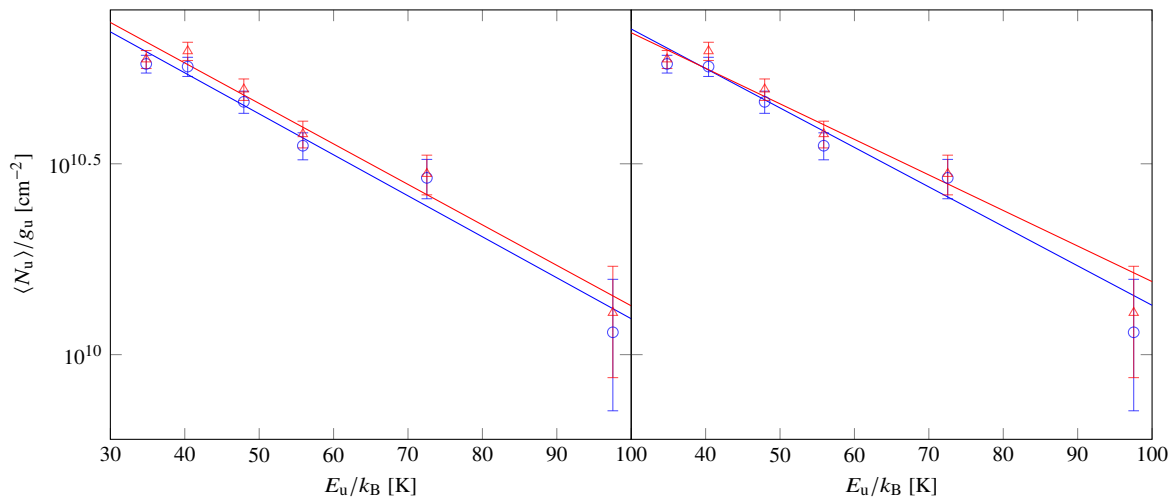
\begin{figure*}
  \centering
  \begin{tikzpicture}
    \begin{semilogyaxis}[
      xlabel={$E_\mathrm{u}/k_\mathrm{B}$ [K]},
      ylabel={$\langle N_\mathrm{u} \rangle/g_\mathrm{u}$ [\si{cm^{-2}}]},
      xmin=30, xmax=100,
      ymin=6e9, ymax=8e10,
      name=rotdiag,
      ]
      \addplot+[blue,mark=o,only marks,
		opacity=0.7,
		error bars/.cd,y dir=both,y explicit] table
      [x index=0,y index=1,y error index=2]
      {rot_ch3oh_AP-H201-X201_2015-01-17.txt};
      \addplot[blue,mark=none] table [x index=0,y index=1]
      {rotdiag_AP-H201-X201_2015-01-17.txt};
      \addplot+[red,mark=triangle,
	                opacity=0.7,
		       	only marks,error bars/.cd,y dir=both,y explicit] table
      [x index=0,y index=1,y error index=2]
      {rot_ch3oh_AP-H201-X202_2015-01-17.txt};
      \addplot[red,mark=none] table [x index=0,y index=1]
      {rotdiag_AP-H201-X202_2015-01-17.txt};
    \end{semilogyaxis}
    \begin{semilogyaxis}[
      xlabel={$E_\mathrm{u}/k_\mathrm{B}$ [K]},
      xmin=30, xmax=100,
      ymin=6e9, ymax=8e10,
      yticklabel=\empty, ylabel=\empty,
      xtick={40,50,60,70,80,90,100},
      at={(rotdiag.east)},anchor=west,
      ]
      \addplot+[blue,mark=o,only marks,
		opacity=0.7,
		error bars/.cd,y dir=both,y explicit] table
      [x index=0,y index=1,y error index=2]
      {rot_ch3oh_AP-H201-X201_2015-01-17.txt};
      \addplot[blue,mark=none] table [x index=0,y index=1]
      {rotdiag_AP-H201-X201_2015-01-18.txt};
      \addplot+[red,mark=triangle,
	                opacity=0.7,
		       	only marks,error bars/.cd,y dir=both,y explicit] table
      [x index=0,y index=1,y error index=2]
      {rot_ch3oh_AP-H201-X202_2015-01-17.txt};
      \addplot[red,mark=none] table [x index=0,y index=1]
      {rotdiag_AP-H201-X202_2015-01-18.txt};
    \end{semilogyaxis}
  \end{tikzpicture}
  \caption{
    Rotation diagram for \ce{CH3OH} lines in comet \lovejoy{}
    including 1$\sigma$ uncertainties from observations obtained on
    January UT 17.012 (left-hand panel) and January UT 18.042 January
    (right-hand panel).
    The natural logarithm of the column density of the upper level
    divided by its degeneracy is plotted against the energy of the upper
    level for each transition.  The red data points show measurements
    with the XFFTS1 backend and the blue data points were obtained
    using XFFTS2, excluding blended lines.
    The solid lines show the best linear fits
    to the lines detected with S/N greater than three
    with an averaged
    rotational temperature of \rottemp{} for January UT 17.012 and
    \rtemp{} for January UT 18.042.
  }
  \label{fig:rotdiagram}
\end{figure*}

In Fig.~\ref{fig:rotdiagram} we show the best-fitting of the rotational
diagram for the \ce{CH3OH} lines observed in comet \lj{} for two
different epochs.  From the relative intensities of the individual
\ce{CH3OH} lines between levels with quantum numbers J = 5--4 we
obtain a rotational temperature of \rottemp{} and a beam-averaged column
density of \column{} for January UT 17.012, and \rtemp{} and \columnb{}
for January UT 18.042.  We note that blended methanol lines were not
included in the analysis.  These values of the rotational temperature
are comparable with those of other Oort-cloud comets observed at similar
heliocentric distances but somewhat lower than other measurements in
comet \lj{} \citep{2016A&A...589A..78B,2017ApJ...836L..25P}.  Since
observations obtained with the Keck and IRAM telescopes have a smaller
field of view than APEX, they cover the collisional region of the coma
which has higher rotational temperature than the outer regions of the
coma encompassed in our observations.

The rotational temperatures derived from \ce{CH3OH} multiplets are
especially useful as they are intermediate between the gas kinetic
temperature in the collisional region in the inner coma and the
fluorescence equilibrium temperature in the outer parts of the coma
\citep[see
e.g.,][]{2004come.book..391B,2013A&A...559A..48D,2016A&A...589A..78B}.
Using the non-LTE \ce{CH3OH} model described in
Section~\ref{sec:modeling}, the inferred rotational temperature can be
compared to the prediction from the model for a gas kinetic temperature
equal to the observed rotational temperature.  Assuming a constant gas kinetic
temperature of \SI{53}{\kelvin}, the rotational temperature derived from
the \ce{CH3OH} level population predicted by the model is
\SI{45}{\kelvin}, in agreement with the measured rotational temperature
within uncertainties.

\subsection{Modelling of line emission}
\label{sec:modeling}

We adopt a radiative transfer method based on the non-LTE Line Modelling
Engine code \citep[\lime{};][]{2010A&A...523A..25B}%
\footnote{The code is available under the GNU General Public License
v3.0 at \url{https://github.com/lime-rt/lime}}
to derive the
production rates that include collisions between neutrals and electrons
and radiation trapping effects \citep[see][and references
therein]{2010A&A...521L..50D}.  The radiative pumping of the fundamental
vibrational levels, which are induced by solar infrared radiation and
subsequently decay to rotational levels in the ground vibrational state,
is calculated using the Comet INfrared Excitation
\citep[\cine{};][]{2017JOSS.2017..182D} code.
We used a one-dimensional spherically symmetric version
of the code with a constant outflow velocity by following the
description outlined in \citet{2004ApJ...615..531B}, which has been used
to model water excitation to interpret cometary observations
from the \herschel{} Space Observatory
\citep[see
e.g.][]{2010A&A...518L.149B,2012A&A...544L..15B,2012A&A...546L...4D,2013ApJ...774L..13O}.

We assume an isotropic gas density profile for parent molecules with
constant outflow velocity for the gas released from the nucleus
\citep{1957BSRSL..43..740H}.  Since the observed area covered by the
APEX beam has a radius of about \SI{10000}{\km} at the comet, the
assumption of isotropic outgassing is reasonable.  The number density of
molecules is by
\begin{equation}
  n_\mathrm{p}(r)= \frac{Q}{4\pi r^2v_\mathrm{exp}}\,
  \exp\left(-\frac{r\beta}{v_\mathrm{exp}}\right),
\label{eq:haser}
\end{equation}
where $Q$ is the molecular production rate, $v_\mathrm{exp}$ is the
expansion velocity and $r$ is the nucleocentric distance.  Volatiles
that sublimate off the surface of the comet can be photodissociated when they
are exposed to solar UV radiation; the photodissociation rate
$\beta$ considers the dissociation and ionization of molecules by the
radiation from the Sun and determines the spatial distribution of the
species.  The values of the photodissociation lifetimes for all the
molecules are obtained from \citet{1994JGR....99.3777C}
except for \ce{NH2CHO} that was extracted from \citet{1976NASSP.393..679J}, and were
multiplied by the factor $\rh^{-2}$, where $\rh$ is the heliocentric
distance at the time of the observations.

To derive the \ce{H2CO} production rate we use a coma density
model for a daughter species based on the Haser model
\citep{2004come.book..523C}
given by
\begin{equation}
  n_\mathrm{d}(r)= \frac{Q}{4\pi r^2v_\mathrm{exp}}\,
  \frac{\beta_\mathrm{p}}{\beta_\mathrm{p} - \beta_\mathrm{d}}\,
  \left( \exp\left(-\frac{r\beta_\mathrm{d}}{v_\mathrm{exp}}\right) -
  \exp\left(-\frac{r\beta_\mathrm{p}}{v_\mathrm{exp}}\right) \right),
\label{eq:haser_daughter}
\end{equation}
where $\beta_\mathrm{p}$ and $\beta_\mathrm{d}$ denote the parent and
daughter photodissociation rates, respectively. We adopt a parent species
photodissociation rate of $\beta_\mathrm{p} = \SI{1e-4}{\per\s}$.
However, the \ce{H2CO} parent source in cometary coma is currently subject to
debate, and the retrieved \ce{H2CO} production rate is very uncertain.

The derived production rate depends on the model input parameters and
collisional excitation rates, as well as on the radiative transfer
method that is used to calculate the level populations and synthetic
spectrum.  We use the molecular data for the observed species available
from the Leiden Atomic and Molecular Database
\citep[LAMDA;][]{2005A&A...432..369S}, including energy levels,
statistical weights and collisional rates. Since LAMDA contains
collisional data from quantum calculations of each molecule with
molecular hydrogen
and the main collisional species in cometary coma is \ce{H2O}, we have
scaled the collision rates in the molecular data files by multiplying
with the ratio of their molecular masses $m_{\ce{H2O}}/m_{\ce{H2}}$
using the \texttt{astroquery} affiliated package of \texttt{astropy}
\citep{astroquery}.  Given the APEX beam size of \SI{~10000}{\kilo\m} at
the comet, the contribution of the radiative pumping of vibrational
levels by solar radiation is not dominant on the production rates, but
there is a noticeable effect on the population levels at distances of
the order of the beam FWHM.

To derive molecular abundances of the observed species relative to
water, the \ce{H2O} production rates were obtained from Odin
observations and measurements of the OH radical maser lines at
\SI{18}{\cm} with the Nan\c{c}ay radio telescope
\citep{2015SciA....115863B,2016A&A...589A..78B}.  During the January
2015 12.8--16.8 observing period, an average water production rate of
\SI{5.0(2)e29}{\mols} was measured.  Based on Nan\c{c}ay and Odin
observations, we adopt a value for $Q_{\ce{H2O}}$ of \SI{6e29}{\mols} for
the period of the APEX observations.  We assumed that the $A$ and $E$
forms of \ce{CH3OH} are equally abundant in our calculation.  As an
estimate of the kinetic gas temperature in the coma we use the value of
the rotational temperature derived in Section~\ref{sec:rottemp} from
multiple \ce{CH3OH} transitions observed simultaneously, \rott{}. We
assume that the kinetic gas temperature and outgassing velocity are
constant throughout the coma in our model.  The outgassing velocity in
the coma is obtained from the half width at half-maximum (HWHM) of a
Gaussian fit to the \ce{CH3OH} transitions detected with high S/N.  The
HWHM was further reduced by \SI{15}{\percent} to take into account the increase in
line width due to thermal Doppler broadening, resulting in a value of
\SI{0.8}{\km\per\s}. This expansion velocity agrees with the value
determined using the IRAM \SI{30}{\m} telescope
\citep{2016A&A...589A..78B}.

Once the level populations are calculated for each grid point in the
cometary model, the synthetic emission line profiles are obtained by ray
tracing along straight lines of sight using \lime{}. Image cubes
are produced for each transition in standard FITS format.  The resulting
line spectrum was obtained by convolution with the APEX beam at each
frequency given in Table~\ref{tab:log}.  The line intensity in the
synthetic spectrum is then compared with the observed integrated
intensities, and the model is varied accordingly using a different value
of the production rate until there is good agreement.

Sensitive upper limits on acetaldehyde and formamide are calculated
by averaging the strongest lines in the wavelength range covered by our
observations, as predicted by our model.  Table~\ref{tab:nh2cho} shows
the chosen emission lines covered by our observations in any of the
receiver settings with either of the XFFTS backends.  All the
considered \ce{CH3CHO} transitions are of the E form.  We assume that
the ratio of A and E forms is also the statistical equilibrium value to
derive the production rate upper limit for \ce{CH3CHO}.   The upper
limits to the averaged line intensities were obtained by integrating the
flux over a width of \SI{1.6}{\km\per\s} derived from the high-S/N
methanol lines. Using a spherically symmetric Haser model, the
resulting 3$\sigma$ upper limits are \SIlist{6e26;e26}{\mols} for
\ce{CH3CHO} and \ce{NH2CHO}, respectively.

\begin{figure*}
  \centering
  \begin{tikzpicture}
    \begin{axis}[
      ybar,
      ymode=log,
      width=\hsize,
      height=9cm,
      log origin=infty,
      enlarge x limits=0.18,
      area legend,
      ylabel={$Q(\ce{X})/Q(\ce{H2O})$ [\si{\percent}]},
      xtick={0,1,2,3},
      xticklabels={\hb{},C/2013 R1,C/2012 F6,C/2014 Q2},
      cycle list name=Dark2-8,
      every axis plot/.append style={fill,draw=none,no markers},
      bar width=8.5pt,
      ymajorgrids=true,
      yminorgrids=true,
      extra x ticks       = {0.5, 1.5, 2.5},
      extra x tick labels = ,
      extra x tick style  = { grid = major },
      ]
      \addplot coordinates{(0,23) (1,7) (2,4) (3,1.9)};
      \addplot coordinates{(0,2.2) (1,2.6) (2,1.6) (3,2.4)};
      \addplot coordinates{(0,0.09) (1,0.12) (2,0.07) (3,0.028)};
	\draw[black,ultra thick,->,>=stealth,xshift=-1.5*10.5pt]
	  (axis cs:2,0.07) -- (axis cs:2,0.005);
      \addplot coordinates{(0,0.025) (1,0.10) (2,0.07) (3,0.047)};
      \addplot coordinates{(0,0.02) (2,0.3) (3,0.12)};
	\draw[black,ultra thick,->,>=stealth,xshift=.5*10.5pt]
	  (axis cs:0,0.02) -- (axis cs:0,0.005);
	\draw[black,ultra thick,->,>=stealth,xshift=.5*10.5pt]
	  (axis cs:2,0.3) -- (axis cs:2,0.005);
      \addplot coordinates{(0,0.10) (1,0.02) (2,0.025) (3,0.009)};
      \addplot coordinates{(0,0.02) (1,0.021) (2,0.0056) (3,0.008)};
      \addplot coordinates{(0,0.05) (1,0.07) (2,0.05)};
	\draw[black,ultra thick,->,>=stealth,xshift=3.5*10.5pt]
	  (axis cs:1,0.07) -- (axis cs:1,0.005);
	\draw[black,ultra thick,->,>=stealth,xshift=3.5*10.5pt]
	  (axis cs:2,0.05) -- (axis cs:2,0.005);
      \legend{
	  \ce{CO},
	  \ce{CH3OH},
	  \ce{HCOOH},
	  \ce{CH3CHO},
	  \ce{C2H5OH},
	  \ce{HNCO},
	  \ce{NH2CHO},
	  \ce{C2H2},
      }
    \end{axis}
  \end{tikzpicture}
  \caption{Observed mixing ratios relative to water
    in comets C/1995 O1 (Hale-Bopp),
    C/2013 R1 (Lovejoy),
    C/2012 F6 (Lemmon) and
    C/2014 Q2 (Lovejoy) from
    this work and from
    \citet{2001Icar..153..162D},
    \citet{2014AJ....147...15P},
    \citet{2014ApJ...791..122P},
    \citet{2014A&A...566L...5B}
    and \citet{2015SciA....115863B}.
    Upper limits are indicated by the black
    vertical arrows.
  }
  \label{fig:com}
\end{figure*}
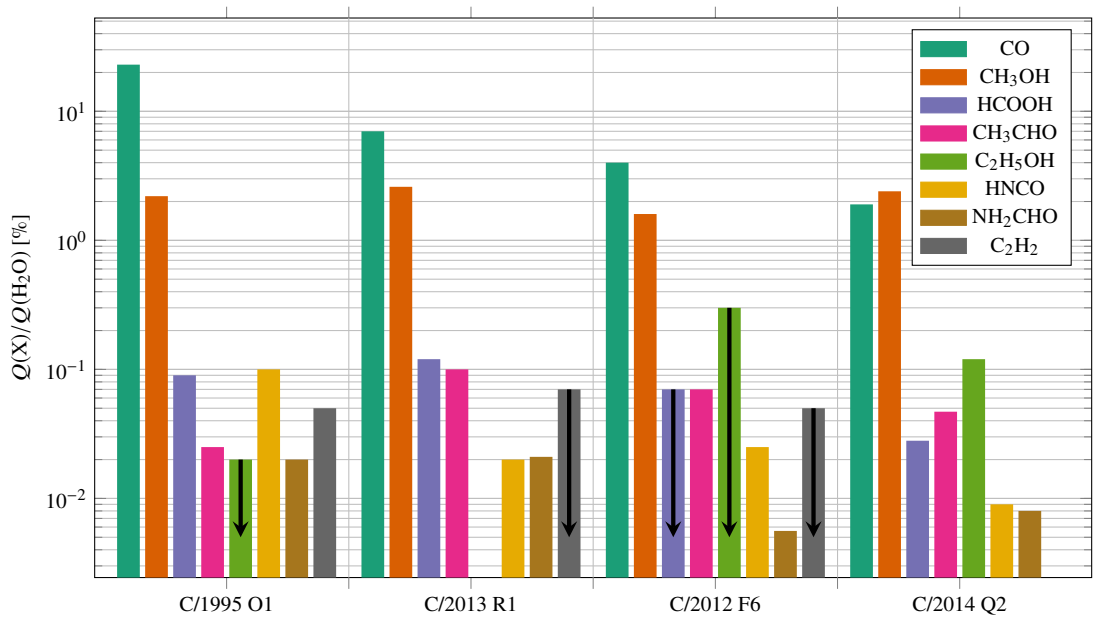

Table~\ref{tab:prodrates} summarizes the production rates
and mixing ratios of the detected volatiles in \lj{} with
respect to \ce{H2O} and a comparison to other measurements using the
IRAM \SI{30}{\m} telescope.  Statistical uncertainties are 1$\sigma$ rms
noise from the integrated intensities.  We confirm that \lj{} is
rather depleted in CO and has low \ce{HCN} and \ce{H2CO} abundances.
The measured production rates are consistent with values derived for
this object from other facilities at radio wavelengths
\citep{2016A&A...589A..78B,2016A&A...588A..72W}.

\begin{table*}
\begin{threeparttable}
  \caption{Production rates in \lovejoy{} with statistical
      uncertainties measured by APEX between 2015 January 16--18.}
  \label{tab:prodrates}
  \begin{tabular}{c
              S[table-format = <1.1(1)e2]
              S[table-format = <1.2]
              S[table-format = 1.3]
              S[table-format = 1.3]
              S[table-format = 2.3]
           }
    \toprule
    Molecule & {$Q$} &
    {$Q/Q_{\ce{H2O}}$\tnote{\emph{a}}} &
    {$Q/Q_{\ce{H2O}}$\tnote{\emph{b}}} &
    \multicolumn{2}{c}{Range $Q/Q_{\ce{H2O}}$\tnote{\emph{c}}}
    \\
    \cmidrule{5-6}
    & {(\si{\mols})} & {(\si{\percent})} & {(\si{\percent})}&
    {Lower (\si{\percent})} & {Upper (\si{\percent})} \\
    \midrule
    \ce{HCN}    & \qhcn   & \hcn   & 0.09  & 0.08  & 0.25  \\
    \ce{CO}     & \qco    & \co    & 1.8   & 2     & 33    \\
    \ce{H2CO}   & \qhco   & \hco   & 0,3   & 0.1   & 1.4   \\
    \ce{CH3OH}  & \qchoh  & \choh  & 2.4   & 0.6   & 6     \\
    \ce{CH3CHO} & \qchcho & \chcho & 0,047 & 0.052 & 0.08  \\
    \ce{NH2CHO} & \qnhcho & \nhcho & 0.008 & 0.012 & 0.021 \\
    \bottomrule
  \end{tabular}
  \begin{tablenotes}
    \item [\emph{a}] Abundances measured by APEX relative to water. An
        average \ce{H2O} production rate of \SI{6.0(6)e29}{\mols} was
        obtained by interpolation of Odin observations during the
        period January 30 to February 3 and Nan\c{c}ay measurements on
        January 12--16 \citep{2015SciA....115863B}.  \ce{CH3CHO} and
        \ce{NH2CHO} abundances are derived 3$\sigma$ upper limits.
	\item [\emph{b}] Relative abundances obtained by the \SI{30}{\m}
        IRAM telescope
        \citep{2015SciA....115863B}.
    \item [\emph{c}] Range of typical abundances measured in a sample of comets.
  \end{tablenotes}
\end{threeparttable}
\end{table*}

\section{Discussion}
\label{sec:discussion}

Our upper limits on \ce{CH3CHO} and \ce{NH2CHO} are consistent with the
mixing ratios measured in \lj{} by \citet{2015SciA....115863B}. These molecules
are amongst a suite of organic molecules found in comets:  \ce{CH3OH}
(methanol), \ce{CH3CHO} (acetaldehyde), \ce{C2H5OH} (ethanol), HCOOH
(formic acid), HNCO (isocyanic acid) and \ce{NH2CHO} (formamide)
that may share a common formation pathway from CO
\citep[e.g.,][]{2011ARA&A..49..471M,2015SSRv..197....9C}.
In this scenario CO molecules are frozen out on cold dust grains and
hydrogen atom additions in CO ice produce the formyl radical from which
larger molecules are built up by subsequent heavy atom additions and
further H additions in the reaction sequences
\citep[e.g.,][]{2008SSRv..138...59C,2009ARA&A..47..427H}:
\begin{align}
\ce{& CO ->[H] HCO ->[H] H2CO ->[H] CH2OH ->[H] CH3OH         \\
    & HCO ->[O] HCOO ->[H] HCOOH                     \\
    & HCO ->[N] HNCO ->[2H] NH2CHO                   \\
    & HCO ->[C] HCCO ->[H] CH2CO ->[2H] CH2CO \notag\\
    & ->[2H] CH3CHO ->[2H] C2H5OH
    \label{reaction:4}}
\end{align}
Most of these reactions have been demonstrated in the laboratory
\citep[e.g.,][]{2015arXiv150702729L}.

A wide range of CO mixing ratios has been measured in comets, with \lj{}
having one of the lowest. By comparing the mixing ratios of complex
organic molecules with their CO content, we can explore the feasibility
of these molecules having had their precometary origin on cold
interstellar/nebular dust grains \citep[e.g.,][]{2008SSRv..138..127D}.
Fig.~\ref{fig:com} summarizes the relative mixing ratios of eight
organic molecules as measured in four comets which displayed a range of
CO mixing ratios.
Mixing ratios in comets \halebopp{}, \rlovejoy{}, \lemmon{} and \lovejoy{}
are from this work and from
the compilations of \citet{2014A&A...566L...5B,2015SciA....115863B},
except for \ce{C2H2} where values are taken from
\citet{2001Icar..153..162D}, \citet{2014AJ....147...15P}, and \citet{2014ApJ...791..122P}.
Fig.~\ref{fig:com}
shows that the \ce{CH3OH} mixing ratio is constant and independent of
the amount of CO present. Although \lj{} has the least CO its methanol
ratio is one of the largest in the sample, with about \SI{56}{\percent} of the
original CO having been converted. This suggests that \ce{CH3OH}
formation is highly efficient, and is probably only limited by the
availability of H atoms on the surface. High CO conversion efficiencies
have been reported in laboratory studies.
The mixing ratios of both HNCO and \ce{NH2CHO} both increase with that
of CO, although the decline of \ce{NH2CHO} is less pronounced. For
\ce{CH3CHO} and \ce{C2H5OH} the trend is puzzling and seems to be
inconsistent with their formation in sequence~\ref{reaction:4}. \rlj{} has the
highest mixing ratios of \ce{CH3CHO} but \lj{}, with the lowest
CO content ($Q(\ce{CO})/Q(\ce{H2O}) = \SI{1.9}{\percent}$), is more
enriched in \ce{CH3CHO} than
\hb{} ($Q(\ce{CO})/Q(\ce{H2O}) = \SI{23}{\percent}$). This suggests that \ce{CH3CHO} and
\ce{C2H5OH} have alternative ice formation processes that are
unconnected to CO. Hydrogenation and O atom addition sequences
starting from \ce{C2H2} are possible \citep[e.g.,][]{2004AdSpR..33...23C}
but is unsupported by the trend of their mixing ratios with the
existing \ce{C2H2} data.  \lj{} does have the lowest HCOOH mixing ratio but
a decline with CO ratio only appears at $Q(\ce{CO})/Q(\ce{H2O}) \approx
\SI{7}{\percent}$.

We conclude that there is some observational support for atom addition
reactions on cold dust being the origin of some of these molecules.
It is not clear why the extreme CO ratio of \hb{} only leads to
modest abundances of complex molecules (apart from HNCO) and why \rlj{} has
the largest abundances. An important caveat to the above analysis is
that we have not considered possible, and essentially unknowable,
variations in the abundances of the heavy atoms (C, O, N) in the gas
where the precometary ices formed. A larger statistical sample of
similar cometary observations will be required to further test these
ideas.

\section{Conclusions}
\label{sec:conclusions}

The chemical composition of the long-period very active comet \lovejoy{}
has been investigated using submillimetre spectroscopic molecular
observations obtained with the \SI{12}{\m} APEX telescope.  We have
presented the results of our data analysis of the observations of
\volatiles{} and derivation of their molecular abundances using a
non-LTE molecular excitation method.  Based on the average of the
strongest lines in the wavelength range covered by our observations, we
have also derived upper limits on the complex molecules acetaldehyde
(\ce{CH3CHO}) and formamide (\ce{NH2CHO}).

We have presented a three-dimensional coma model that allows a detailed
interpretation of cometary observations and computation of the
production rates from the observed line transition intensities. We
calculate the molecular excitation in the coma using the \lime{} code by
\citet{2010A&A...523A..25B}.  This model includes collisional excitation
of the rotational levels by water, spontaneous and induced emission, and
radiation trapping effects.  The radiative pumping of the fundamental
vibrational levels induced by solar infrared radiation which
subsequently decay to rotational levels in the ground vibrational state
is calculated using the \cine{} code \citep{2017ascl.soft08002D}.
Although we have used an isotropic outgassing geometry to interpret
our observations, the capability of the code to handle multidimensional
geometries, e.g., collimated jets or distributed sources in the coma,
will be useful for the interpretation of future comet observations
carried out with ALMA.

Based on a spherically symmetric Haser model with constant outflow
velocity, and assuming a constant kinetic gas temperature
from observations of multiple \ce{CH3OH} transitions in comet \lj{}, we
derive molecular production rates for \volatiles{}, and upper limits for
\ulimits{}. The methanol rotational temperature of \rottemp{} derived
from multiple lines using the rotational diagram technique is lower and
consistent within the accuracy of the detections with the value of
\tiram{} for \ce{CH3OH} determined by \citet{2016A&A...589A..78B}.
Our measurement is also lower than the \ce{H2O} rotational temperature
of \tkeck{} retrieved by \citet{2017ApJ...836L..25P}.  This
inconsistency in the derived rotational temperatures between infrared and
radio measurements can be reconciled taking into account that
observations with small beam size (infrared) sample a hotter region of
the coma while observations with a larger beam (radio) include cooler
material and thus exhibit a lower rotational temperature.
We discussed the possible connection between the low CO mixing ratios
and the relative abundances of complex molecules in \lj{} in comparison
with other comets. Further observations of future cometary  apparitions,
similar to those presented here, will provide important information on
chemical formation pathways and physical conditions in the protosolar
molecular cloud core and/or the protosolar nebula
\citep[e.g.,][]{2015SSRv..197....5M,2015SSRv..197..151W}.

\section*{Acknowledgements}

This work was supported by NASA's Planetary Astronomy Program.
This publication is based on data acquired with the Atacama Pathfinder
Experiment (APEX) telescope. APEX is a collaboration between the
Max-Planck-Institut f\"ur Radioastronomie, the European Southern
Observatory, and the Onsala Space Observatory.
The APEX observations were conducted under the observing proposal
O-094.F-9307(A).
The authors would like to thank the referee, J.~Crovisier, for giving
valuable comments.
We gratefully acknowledge the support of the APEX
staff, in particular to Michael Olberg and Dirk Muders, for their
assistance with data analysis.

\bibliographystyle{mnras}
\bibliography{ads,q2}

\bsp 
\label{lastpage}

\end{document}